\documentclass[%
reprint,
superscriptaddress,
 amsmath,amssymb,
prl,
longbibliography,
]{revtex4-1}

\usepackage{romannum}
\usepackage{graphicx}
\usepackage{dcolumn}

\usepackage{siunitx}
\usepackage{hyperref}

\usepackage{xcolor}
\definecolor{myred}{RGB}{255,3,0}
\definecolor{myblue}{RGB}{31,60,255}
\definecolor{myblue2}{RGB}{34,123,255}
\definecolor{myorange}{RGB}{226,48,0}
\definecolor{mygreen}{RGB}{21,100,0}

\usepackage{tikz}
\usetikzlibrary{shapes.geometric,shapes.symbols}

\begin{document}

\title{\large{Deforming ice with drops}}

\author{Duco van Buuren}
\affiliation{Physics of Fluids group, Max Planck Center Twente for Complex Fluid Dynamics, Department of Science and Technology, Mesa+ Institute and J. M. Burgers Center for Fluid Dynamics, University of Twente, P.O. Box 217, 7500 AE Enschede, The Netherlands}

\author{Pallav Kant}
\email[]{kantpallav88@gmail.com}
\affiliation{School of Engineering, University of Manchester, M13 9PL, United Kingdom}

\author{Jochem G. Meijer}
\affiliation{Physics of Fluids group, Max Planck Center Twente for Complex Fluid Dynamics, Department of Science and Technology, Mesa+ Institute and J. M. Burgers Center for Fluid Dynamics, University of Twente, P.O. Box 217, 7500 AE Enschede, The Netherlands}

\author{Christian Diddens}%
\affiliation{Physics of Fluids group, Max Planck Center Twente for Complex Fluid Dynamics, Department of Science and Technology, Mesa+ Institute and J. M. Burgers Center for Fluid Dynamics, University of Twente, P.O. Box 217, 7500 AE Enschede, The Netherlands}

\author{Detlef Lohse}%
\email[]{d.lohse@utwente.nl}
\affiliation{Physics of Fluids group, Max Planck Center Twente for Complex Fluid Dynamics, Department of Science and Technology, Mesa+ Institute and J. M. Burgers Center for Fluid Dynamics, University of Twente, P.O. Box 217, 7500 AE Enschede, The Netherlands}
\affiliation{Max Planck Institute for Dynamics and Self-Organization, Am Fassberg 17, 37077 G\"ottingen, Germany}

\begin{abstract}
A uniform solidification front undergoes non-trivial deformations when encountering an insoluble dispersed particle in a melt. For solid particles, the overall deformation characteristics are primarily dictated by heat transfer between the particle and the surrounding, remaining unaffected by the rate of approach of the solidification front. In this Letter we show that, conversely, when interacting with a droplet or a bubble, the deformation behavior exhibits entirely different and unexpected behavior. It arises from an interfacial dynamics which is specific to particles with free interfaces, namely thermal Marangoni forces. Our study employs a combination of experiments, theory, and numerical simulations to investigate the interaction between the droplet and the freezing front and unveils its surprising behavior.
In particular, we quantitatively understand the dependence of the front deformation $\Delta$ on the front propagation velocity, which, for large front velocities, can even revert from attraction ($\Delta < 0$) to repulsion ($\Delta > 0$). 
\end{abstract}

\date{\today}

\maketitle

\begin{figure}[b!]
 \centerline{\includegraphics[width=0.8\columnwidth]{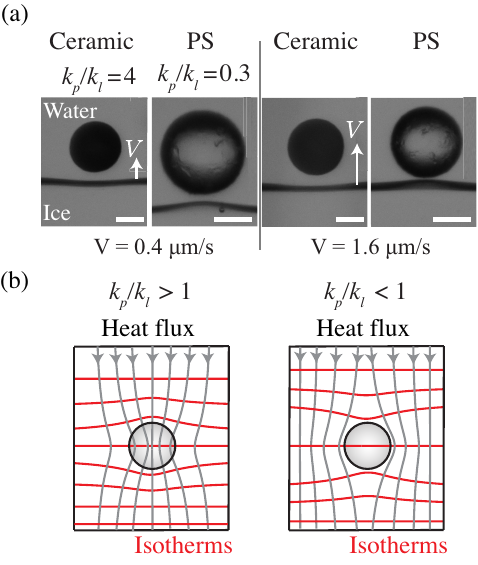}}
\caption{\textbf{Deformation of the ice front by a solid particle}. (a) Shape of the solidification front approaching a ceramic and polystyrene (PS) particle at different advancing velocities $V$. Dependent on the thermal conductivity mismatch, the front is either deflected away (left, $k_p/k_{l} = 4$) or towards (right, $k_p/k_{l} = 0.3$) the particle. The direction of the deflection is independent on $V$. The scale-bars are $\SI{50}{\micro \meter}$.
  (b) Schematic of the origin of the front deflection for particles more (left) or less (right) conductive than the surrounding liquid when an uniform temperature gradient is applied. The solidification front corresponds to the isotherm (red lines) that matches the melting temperature.}
\label{fig:1}
\end{figure}

During the solidification of an aqueous suspension, the dispersed objects (particles, drops, bubbles, etc.) either become an integral part of the solidified bulk or get rejected by it \citep{shangguan1992analytical,rempel2001particle, park2006encapsulation,dedovets2018five,tyagi2020objects,meijer2023frozen}.  
The prevailing outcome therefore controls the micro-structure of the solidified material and thus its functional properties.
Crucially, the conditions dictating the engulfment or rejection of a dispersed particle rely on a delicate interaction between the particle/drop/bubble and the moving solidification front, where the relevant physical processes act across various length scales. 
At length scales much smaller than the size of the dispersed particle, this interaction is controlled by a combination of repulsive van der Waals interactions between the dispersed particle and the freezing interface and fluid flow within thin premelted films enveloping the particle \cite{wettlaufer2006premelting, rempel2001particle, meijer2023thin, meijer2023frozen}. 
In contrast, at the particle-scale, it is the heat exchange between the particle and the surrounding liquid, the interfacial energy of the solidification interface, and the density change caused by the liquid–solid phase transition that controls the dynamics.
Previous investigations have revealed that, as a solidification front approaches a dispersed particle, it experiences deformation in a non-trivial manner \cite{shangguan1992analytical,tyagi2020objects}.
It is largely accepted that this deformation of the solidification front is mainly controlled by the mismatch in thermal conductivity between that of the particle $k_p$ and the surrounding liquid $k_l$ which, in turn, dictates the heat transfer between the two \cite{shangguan1992analytical}.
Most importantly, in case of solid particles, the rate of approach of the solidification front has no effect on the deformation of the solidification front (see Fig.\,\ref{fig:1}(a)) \citep{shangguan1992analytical, tyagi2020objects}.

In contrast, our Letter introduces a remarkable phenomena occurring when an immiscible oil droplet immersed in water encounters an advancing solidification front:
We demonstrate that, while the deformation of the freezing front at low advancing velocities conforms to the expected behaviour for a given thermal conductivity mismatch, \textit{i.e.}, deflection towards the drop since $k_p/k_l < 1$ (see Fig.\,\ref{fig:1}(b)),  an unexpected \textit{reversal} of the ice front deflection occurs at faster freezing rates. 
We employ a comprehensive approach, integrating controlled experiments, theoretical insights, and numerical simulations to elucidate the origin of the observed reversal. Our analysis reveals that this reversal is attributed to a self-induced flow along the droplet interface, driven by temperature-induced gradients in surface tension - a phenomenon known as the thermal Marangoni effect. 
Such a fluid flow transports warmer water toward the colder ice front, leading to localized increase of the temperature field and thus delayed freezing.
Since the shape of the interface plays a crucial role in determining the critical engulfment velocity, beyond which particles become trapped in the solidifying material \cite{park2006encapsulation,meijer2023frozen}, our findings contribute novel insights also to this aspect. 
Consequently, this study holds significance not only in terms of fundamental understanding but also for practical applications.
Note that the deformation of the freezing interface is rather distinct - both in underlying physical origin and in phenomenology - from the deformation of the oil droplet during engulfment into the solidifying bulk, which we presented in our earlier work Ref.\,\cite{meijer2023thin}.

The experimental setup used to investigate the deformation of the advancing solidification front by dispersed droplets is similar to the one used in our previous investigations \cite{meijer2023thin, meijer2023frozen,meijer2023freezing}.
For brevity, here, we highlight only the important details for the setup and refer to Ref.~\cite{meijer2023thin} for detailed descriptions.
We investigate the interaction between an advancing water-ice freezing front and silicone oil droplets in a horizontal Hele-Shaw cell of thickness 200\,$\mu$m, filled with an oil-in-water emulsion.
The emulsion is prepared using a microfluidic setup, as explained in Ref.~\cite{van2021feedback}.
To ensure the stability of the emulsion, a surfactant TWEEN80 (Sigma-Aldrich, Germany) is added to the water (0.5 vol\% initially). 
Since surfactants are known to affect freezing of complex liquids \cite{dedovets2018five,kao2009particle,tyagi2022solute}, the emulsion sample is diluted after fabrication to achieve low surfactant concentration (0.01 vol\%).
As a consequence, any such solutal effects are irrelevant for our experiments.
In each experiment, the Hele-Shaw cell is translated at constant velocity $V$ through a fixed temperature gradient $G = \nabla T$ using a linear actuator (Physik Instrumente, M-230.25) such that the position of the moving solidification front remains fixed in the lab frame of reference. 
The temperature gradient $G \sim 10\,\mathrm{K\,cm^{-1}}$ is applied over the entire length of the Hele-Shaw cell by placing it over a copper block with cold and warm ends, maintained at fixed temperatures using Peltier elements.
The applied thermal gradient sets the ice growth rate and also thus the approach velocity $V$ of a droplet to the front.
Features of interest along the moving freezing front are recorded in top-view through a camera connected to a long working distance lens (Thorlabs, MVL12X12Z), and diffused back lighting. 

\begin{figure}
  \centerline{\includegraphics[width=0.8\columnwidth]{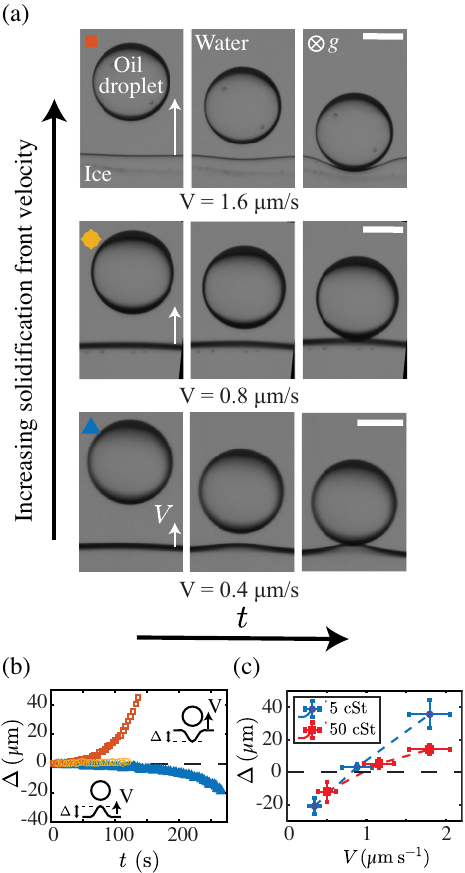}}
  \caption{\textbf{Deformation of the ice front by a droplet}. (a) Shape of the solidification front as it approaches a silicone oil droplet ($\SI{5}{cSt}$) of size $R \approx \SI{90}{\micro \meter}$ at three different advancing velocities ($V = \SI{0.4}{\micro \meter \per \second}, \SI{0.9}{\micro \meter \per \second},  \SI{1.6}{\micro \meter \per \second}$, see Suppl. Movies 1-3).  Although the mismatch in thermal conductivity remains unaltered ($k_p/k_{l} = 0.2$), a reversal of the solidification front deflection is observed as the front approaches faster. The scale-bars are $\SI{100}{\micro \meter}$.  (b) Furthest deflection $\Delta$ of the front, with respect to the planar far field, as function of time for the three different advancing velocities (different symbols) corresponding to (a). (c) Deflection of the front when the drop touches the ice (right panels in (a)) as a function of the advancing velocity $V$.  When increasing the viscosity of the drop from low (blue circles, $\SI{5}{cSt}$) to high (red squares, $\SI{50}{cSt}$), the reversal of the solidification front deflection at larger velocities is less pronounced. }
\label{fig:2}
\end{figure}

The sequences of images in Fig.\,\ref{fig:2}(a) illustrate distinct deformation behaviors of a planar solidification front as it encounters a dispersed oil droplet ($R\sim \SI{90}{\micro \meter}$) at different velocities. 
At a low advancing velocity ($V = \SI{0.4}{\micro \meter \per \second}$), as shown in Fig.\,\ref{fig:2}(a), the initially planar solidification interface gradually deforms towards the droplet with its approach, creating an impression that the droplet is exerting local pull on the solidification front.
This deformation behavior, consistent with previous studies, is expected in the presence of a particle with lower thermal conductivity than its surroundings, due to the thermal conductivity of the oil droplet being less than that of water.
In this scenario, the heat flow chooses the path of least resistance, thereby here avoiding the particle itself. 
This, in turn, causes the otherwise uniform isotherms to deflect as illustrated in Fig.\,\ref{fig:1}(b).
Correspondingly, the solidification front simply follows the isotherm, which matches the melting temperature of water \citep{shangguan1992analytical,rempel2001particle, park2006encapsulation,meijer2023frozen}.
Correspondingly, for small P\'{e}clet numbers $\mathrm{Pe} = V\,R/\kappa_l \ll 1$, with $\kappa_l$ the thermal diffusivity of the liquid, the distortion of the isotherms surrounding a particle placed in a uniform thermal gradient can be calculated by solving the heat equation $\nabla^2 T_B = 0$ in all phases.
The analytical solution of the heat equation with appropriate heat flux boundary conditions at the particle surface and the simplifying assumption that both the solidified bulk and the melt have the same thermal conductivity yield \cite{chernov1977, shangguan1992analytical}
\begin{equation}
T_B(\mathbf{x}) = T_{\mathrm{ref}} + \mathbf{G} \cdot (\mathbf{x}-\mathbf{x}_0)  + k_eR^3  \, \frac{ \mathbf{G} \cdot (\mathbf{x}-\mathbf{x}_0)  }{\vert \mathbf{x} - \mathbf{x}_0 \vert^3}, 
\label{Eq:ThermOuter}
\end{equation}
where $k_e = (k_l-k_p)/(2k_l+k_p)$, with $-1<k_e<1/2$, is the dimensionless parameter that quantifies the mismatch in the thermal conductivities and $T_{\mathrm{ref}}$ the reference temperature at the center of the drop, located at $\mathbf{x}_0$.
When the particle and liquid have different thermal conductivities,  $k_e$ is either negative ($k_p/k_l > 1$) or positive ($k_p/k_l < 1$) and the interface deflects away from or towards the particle, respectively (see Fig.\,\ref{fig:1}).
Note that despite the simplified approach of this model, it produces result that closely match with experimental observations \cite{shangguan1992analytical}.

Since the above model description excludes any dynamic effects arising due to heat transfer due to convection and viscous dissipation, it only follows that for a given thermal conductivity mismatch between the particle and the surrounding liquid, the deformation of the solidification front remains unaffected by the rate of engulfment.
While this holds for solid particles (see Fig.\,\ref{fig:1}(a)), unexpected deformation behaviors of the solidification front are observed at higher solidification rates when encountering a droplet.
We find that as the solidification rate is increased to a higher value $V = \SI{0.9}{\micro \meter \per \second}$, the solidification front remains planar as it approaches the dispersed droplet. 
A subsequent further increase in the rate of solidification to $V = \SI{1.6}{\micro \meter \per \second}$ even leads to the \textit{reversal} of the deformation of the solidification front, giving an impression as if the droplet is indenting the solid ice.

\begin{figure}[t!]
  \centerline{\includegraphics[width=0.8\columnwidth]{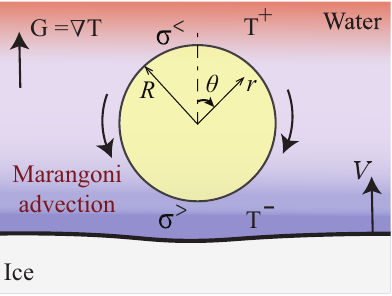}}
  \caption{\textbf{Mechanism of the reversal of the ice front deflection}. Schematic of the thermal Marangoni advection induced by a gradient in surface tension over the interface of the drop $\sigma(T)$, caused by the applied thermal gradient $G$. Warmer water at the top is advected towards the colder ice front, locally increasing the temperature field, leading to enhanced melting in front of the droplet.}
\label{fig:3}
\end{figure}

The physical mechanism responsible for the strange deformation behaviour of the solidification front at faster freezing rates is intimately connected to the interfacial dynamics along the droplet's interface.
Due to the temperature dependence of the liquid's surface tension, it is expected that a droplet in a thermal gradient experiences differential surface tension along its interface.
Accordingly, in our experiments, the part of the droplet closest to the solidification front experiences a higher surface tension than the rest of the surface, given that $\mathrm{d}\sigma/\mathrm{d}T$ is negative for silicone oil.
This temperature-induced difference in surface tension along the droplet interface induces an interfacial flow, drawing liquid from the warmer end to the colder end, which is known as thermal Marangoni flow (see Fig.\,\ref{fig:3}).
This fluid flow surrounding the droplet disturbs the local thermal equilibrium and distorts the base temperature field given by Eq.\,(\ref{Eq:ThermOuter}).

\begin{figure*}[t!]
  \centerline{\includegraphics[width=0.7\textwidth]{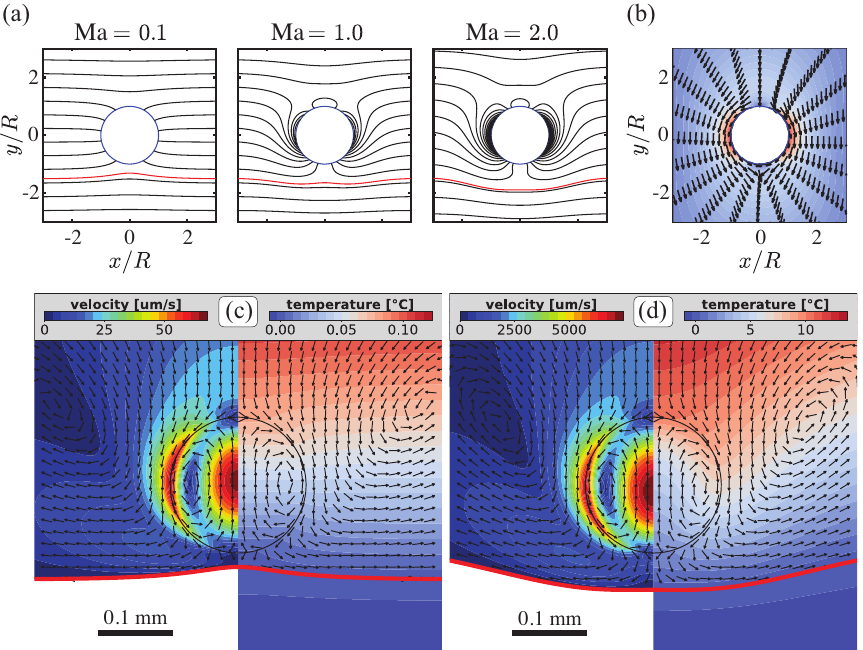}}
  \caption{\textbf{Isotherm perturbation due to thermal Marangoni advection}. (a) Isotherms of the perturbed temperature field around a drop with increasing Marangoni number for $k_p/k_{l} = 0.2$ according to our perturbation theory. The red lines correspond to the isotherms where $T(\mathbf{x}) = T_m$, \textit{i.e.}, the melting temperature of water. (b) Normalised self-induced velocity field around the drop due the thermal Marangoni effect \cite{young1959motion}. The flow is strongest at the equator of the drop (red color).  Arrows only indicate the flow direction. (c)-(d) Numerical simulations of a solid-liquid interface approaching an oil droplet experiencing a thermal gradient of (c) $G = \SI{1e2}{\kelvin \per \meter}$ (implying $Ma = 9$) and (d) $G = \SI{1e4}{\kelvin \per \meter}$ (implying $Ma = 0.07$), respectively. The results qualitatively agree with our experimental observations (see Fig.\,\ref{fig:2}(a)).}
\label{fig:4}
\end{figure*}

To account for this change in the temperature field due to the thermal Marangoni flow around the droplet, we develop a simplified model.
We assume that the droplet is placed in an external thermal gradient without any boundary effects.
For such a case, the analytical description of the flow-field generated by the thermal Marangoni effect was originally derived in Refs.\,\cite{rybczynski1911uber,young1959motion}.
Following the same analysis in our case yields a polar velocity of the fluid flow $u_\theta$ outside the drop as (see Suppl. Mat. Sec. I):
\begin{equation}
u_\theta(r,\theta) = - \frac{1}{2} \frac{1}{\mu+\mu^\prime} \left(\frac{R}{r}+\frac{R^3}{r^3}\right) \frac{\mathrm{d} \sigma}{\mathrm{d} T} \frac{k_l}{2 k_l + k_p} GR \sin\theta,
\label{eq:youngUtheta}
\end{equation}
where $\mu$ and $\mu'$ are the viscosity of water and oil, respectively,  and $\mathrm{d}\sigma/\mathrm{d}T< 0$ quantifies the temperature dependence of the surface tension.
We evaluate Eq.\,(\ref{eq:youngUtheta}) at the equator of the drop and define the thermal Marangoni velocity as
\begin{equation}
V_M \equiv  u_{\theta}\vert_{r = R, \theta = \pi/2} =   - \frac{1}{\mu + \mu'} \frac{\mathrm{d}\sigma}{\mathrm{d}T} \frac{k_l}{2 k_l + k_p}  G R.
\label{Eq:VM}
\end{equation}
From Eq.\,(\ref{Eq:VM}) we see that the flow at the interface of the drop, characterised by $V_M$, is proportional to the size of the drop (not varied in experiments), the strength of the applied thermal gradient $G$, and inversely proportional to $\mu + \mu'$, the sum of the viscosities of water and oil.

To assess the influence of the thermal-Marangoni flow on the local heat transfer near the droplet, we solve the linearized energy equation.
The base state is assumed to be established by heat-transfer via purely conductive effects and is given as $\mathbf{u}_B = 0$ and $\nabla^2\mathbf{T}_B =  0$, where $\mathbf{T}_B$ is given by Eq.\,(\ref{Eq:ThermOuter}). 
This base state is superimposed by the thermal field due to the thermal-Marangoni flow as $T(\mathbf{x}) = T_B(\mathbf{x}) + T_1(\mathbf{x})$ and  $\mathbf{u} =  \mathbf{u}_B +\mathbf{u}_1$, with $\mathbf{u}_1 = [u_r, u_{\theta}]$ (see Suppl. Mat. Sec. I).
Accordingly, the linearized, non-dimensionalized energy equation then reads:
\begin{equation}
Ma \, \tilde{\mathbf{u}}_{1} \cdot \nabla \tilde{T}_B = \nabla^2 \tilde{T}_1,
\label{Eq:ThermConvection_Normalised}
\end{equation}
with tildes denoting dimensionless quantities.
Here, we define the Marangoni number $Ma \equiv V_M R/\kappa$ as the ratio of advective over diffusive thermal transport and it is the same as the P\'{e}clet number introduced earlier.
Typically, here we have $Ma \sim \mathcal{O}(1)$.
All lengths are non-dimensionalized by the droplet radius $R$, velocities by $V_M$, and temperatures as $\tilde{T} = T - T_\mathrm{cold}/(T_\mathrm{m} - T_\mathrm{cold})$, where $T_\mathrm{cold}$ and $T_\mathrm{m}$ are temperatures at the cold end of the experimental cell and the melting point of water, respectively.
We solve Eq.\,(\ref{Eq:ThermConvection_Normalised}) numerically to determine $T_1(\mathbf{x})$ and hence $T(\mathbf{x})$ (see Fig.\,\ref{fig:4}(a)).
The Marangoni number $Ma$ is the natural control parameter dictating the strength of the velocity field around the drop (see Fig.\,\ref{fig:4}(b)) and hence the extent of changes in the local thermal field.
Considering that the freezing interface follows the isotherm matching the melting temperature of water (red lines in Fig.\,\ref{fig:4}(a)), we qualitatively capture the same deformation behaviour as seen in the experiments (see Fig.\,\ref{fig:2}(a)).
In line with Eq.\,(\ref{Eq:VM}), by varying the temperature gradient $G$, and hence the advancing velocity of the front, or by changing the viscosity of the oil (see Fig.\,\ref{fig:2}(c)), we vary the strength of the flow around the drop in our experiments, and thus the extend of the ice deformation.

The simplified approach discussed earlier offers a qualitative depiction of the unique deformation behaviors of the solidification front and its relationship with the applied thermal gradient. 
However, its precision in faithfully simulating the influence of the solidification front on the flow field around the droplet, and thus the overall deformation, is limited. 
Consequently, here we have also introduced a detailed numerical approach that systematically captures all interactions (refer to Supplementary Material, Section II). 
The results obtained from finite element simulations (Fig 4(c,d)) quantitatively agree with the experimental observations.

In summary, our integrated approach involving experiments, theoretical analysis, and numerical simulations disentangles various effects occurring when dispersed droplets in a molten medium interact with an advancing freezing front.
Our findings reveal that, contingent on the freezing rate, a droplet can induce unexpected deformations in the freezing interface, in both directions.
This phenomenon is attributed to the interfacial flow generated by the thermal Marangoni effect, leading to alterations in the temperature field and, consequently, the shape of the solid-liquid interface in its proximity.
Our findings contribute to a better understanding of the mechanism that govern particle rejection or engulfment during the solidification of a complex medium, which is crucial to step towards designing advanced material with functional properties \cite{deville2009metastable}.

\section*{Acknowledgements}
The authors thank Gert-Wim Bruggert and Martin Bos for the technical support. The authors acknowledge the funding by Max Planck Center Twente and the Balzan Foundation.
\bibliography{References}

\end{document}